\documentclass[10pt,showpacs]{revtex4}
\usepackage{amsfonts}
\usepackage{units,hyperref,fancyhdr}
\usepackage{bbold,euler,newcent}
\usepackage{graphicx}
\usepackage{dcolumn}
\usepackage{bm,mathtools,esvect}
\usepackage{slashed}
\begin{document}

\title{\LARGE\sc Quaternionic Electrodynamics\\ \vspace{5mm}}
\author{\sf\large SERGIO GIARDINO}
\email{sergio.giardino@ufrgs.br}
\affiliation{\vspace{3mm} Departamento de Matem\'atica Pura e Aplicada, Universidade Federal do Rio Grande do Sul (UFRGS)\\
Avenida Bento Gon\c calves 9500, Caixa Postal 15080, 91501-970  Porto Alegre, RS, Brazil}

\begin{abstract}
\noindent 
We develop a quaternionic electrodynamics and show that it naturally supports the existence of magnetic monopoles.
We obtained the field equations, the continuity equation, the electrodynamic force law, the Poynting vector, the energy conservation and the stress-energy tensor. 
The formalism also enabled us to generalize the Dirac monopole and the charge quantization rule.
\end{abstract}

\maketitle
\tableofcontents
\section{\sc Introduction \label{I}}

Much of the mathematical non-commutativity widespread in physics may be divided into two areas. The geometric non-commtutativity comprises the local, short range non-commutativity  whose recent
development is mainly due to Allain Connes. We do not tackle this non-commutativity here,
and we only quote \cite{Konechny:2000dp} as a nice introduction to the subject. Conversely, the purely algebraic non-commutativity is global and 
transfers its proper  symmetry to the model that describes a physical system. In this article, the quaternion algebra will transfer its symmetrical properties to  a classic electrodynamic theory
buit over them. 
However, before presenting such a construction we remember that quaternions ($\mathbb{H}$) are hyper-complex numbers endowed with three anti-commutative imaginary units, named $i,\,j$ and $k$, and thus generalizes 
the complex numbers ($\mathbb{C}$). Good introductions to quaternions and their applications in physics may be found in \cite{Anderson:1992wj,Ward:1997qcn,PRG,Rocha:2013qtt}. In this article,
we only mention that $q\in\mathbb{H}$ is such that
\begin{equation}\label{i01}
 q=x_0 + x_1 i + x_2 j + x_3 k, \qquad\mbox{for}\qquad x_0,\,x_1,\,x_2,\,x_3\in\mathbb{R}\qquad\mbox{and}\qquad i^2=j^2=k^2=-1.
\end{equation}
The quaternionic non-commutativity relies on the imaginary units anti-commutativity, so that $\,ij=-ji\,$, by way of example. Alternatively to (\ref{i01}), quaternions can be expressed in symplectic notation, where $\,q=z+\zeta j\,$ for complex $z$ and $\zeta$. We emphasize that the results presented in this article are written in terms of real quaternions (\ref{i01}), and not in the complexified generalization of real quaternions, the biquaternions \cite{Ward:1997qcn}.

There exists an old interest to apply various complex structures to electromagnetism, and we mention applications with 
complex numbers \cite{Kryuchkov:2015xga,Rajantie:2012xg}, quaternions \cite{Winans:1977jgw,Redkov:2011kex,Yefremov:2014jga,Bolokhov:2017ndw,Chanyal:2017sjj,Kansu:2020atu,Chanyal:2019zse,Marques-Bonham:2020fmw}, bi-quaternions 
\cite{Majernik:1999rad,Grudsky:2004gkk, Bisht:2007ju,Weng:2015qoa,Hong:2019azp,Kansu:2019xws}, hyperbolic quaternions \cite{Demir:2010zz}, octonions \cite{Gogberashvili:2005xb,Chanyal:2009xs,Tolan:2013boa,Chanyal:2014edm},  
hyperbolic octonions \cite{Chanyal:2010sz,Marques-Bonham:2020fmw} and sedenions \cite{Mironov:2018bgx}. 

In this article, a novel hyper-complex formulation of electrodynamics is motivated by
a quaternionic version of the quantum Lorentz force recently obtained \cite{Giardino:2019xwm}. The starting point is the quantum quaternionic momentum operator
$\,\bm\Pi\,$ \cite{Giardino:2016abe, Giardino:2018lem}, that acts on a quaternionic wave function $\Psi$ such as
\begin{equation}\label{i1}
\bm\Pi\Psi=-\hbar\big(\bm\nabla-\bm{\mathcal A}\big)\Psi i,\qquad\mbox{where}\qquad \bm{\mathcal{A}}=\bm\alpha i+\bm\beta j
\end{equation}
is the quaternionic vector potential, which comprises a real vector potential $\bm\alpha$ and a complex vector potential $\bm\beta$.  The usual generalized linear momentum operator of complex quantum mechanics ($\mathbb C$QM) is recovered within the limit where $\,\bm\beta=\bm 0.\,$ Therefore, $\,\bm{\mathcal A}\,$ is a pure imaginary quaternionic vector, where each spatial component comprises a quaternionic number.
 The physical meaning of the vectors $\,\bm\alpha\,$ and $\,\bm\beta\,$ will be further explained 
in Section \ref{M}. In the same token as the vector potential is a joining element between the quantum description of the Lorentz force and the classical formulation of electrodynamics, the quaternionic vector potential $\,\bm{\mathcal A}\,$ predicted by the generalized moment (\ref{i1}) indicates the existence of a certain classical quaternionic electrodynamics.  The  vector potential was further shown to generate the magnetic field \cite{Giardino:2019xwm}, namely
\begin{equation}\label{i2}
 \bm{\mathcal B}=\bm{\nabla\times\mathcal A}\,+\,\bm{\mathcal{A}\times\mathcal{A}}\qquad\mbox{for}\qquad\bm{\mathcal{A}\times\mathcal{A}}\,=\,\bm{\bar\beta\times\beta}+2i\,\bm{\alpha\times\beta}\,j,
\end{equation}
 where we emphasize that $\,\bm\alpha\,$ is a real vector and $\,\bm\beta\,$ is a complex vector whose complex conjugate is  $\bm{\bar\beta}$. It is 
immediately seen that $\bm{\mathcal{B}}$ is also pure imaginary, hence their quaternionic conjugate satisfies $\,\overline{\bm{\mathcal B}}=-\bm{\mathcal B}.\,$ We also clarify that the conjugate means the conjugate of the vector components. The vector product between quaternionic vectors is defined in equations (\ref{i02}-\ref{i03}), later in this section.  More importantly, the quaternionic nature of $\,\bm{\mathcal A}\,$ enables the existence of the self-interacting term $\,\bm{\mathcal{A}\times\mathcal{A}}\,$ in (\ref{i2}), a contribution that is absent in the real formulation of electrodynamics. We name this self-interacting electrodynamic theory as quaternionic electrodynamics ($\mathbb H$ED). It is also important to notice  the existence of $\mathbb H$ED as a consistency condition for quaternionic quantum mechanics ($\mathbb H$QM) \cite{Giardino:2018lem,Giardino:2018rhs}. Without it, the quantum Lorentz force does not make sense. Compared to the previous quaternionic formulations \cite{Winans:1977jgw,Redkov:2011kex,Yefremov:2014jga,Bolokhov:2017ndw,Chanyal:2017sjj,Kansu:2020atu}, the quaternionic  formulation of electrodynamics presented in this article is novel because it uses quaternionic vectors, and not quaternionic numbers;  more importantly, the theory is novel because of their self-interacting physical character given in (\ref{i2}). 

The self-interaction component $\,\bm{\mathcal{A}\times\mathcal{A}}\,$ implies that $\,\bm{\mathcal{B}}\,$ is not necessarily divergenceless, thus raising the hypothesis 
of magnetic monopoles solutions within quaternionic quantum mechanics ($\mathbb{H}$QM). However, 
quantum solutions must have a classical limit, and classical electromagnetism is a  theory of real functions. Therefore, we develop in this article a classical theory of quaternion valued electromagnetic 
fields to  serve as the classical limit for $\mathbb{H}$QM, and we study several magnetic monopole solutions as immediate examples. 
Irrespective of this role as the classical limit of $\mathbb{H}$QM, this quaternionic classical
electromagnetism seems  interesting enough to be independently studied. The  vector product that depicts the self-interaction of the vector potential, and the magnetic monopole term
in (\ref{i2}) are both dependent on the quaternionic structure, and are in fact invisible without this algebraic formalism. Thus, the electromagnetic theory presented here is an example of how
a mathematical structure can be useful to describe new physics.

We notice that magnetic monopole descriptions in non-commutative terms are speculated in various models, as
in non-abelian quantum field theories \cite{Bais:2004sc,Rajantie:2012xh}, in the
quaternionic description of  hyperbolic monopoles in Yang-Mills theories \cite{Atiyah:1979gym,Cockburn:2014aya},
in non-commutative quantum mechanics \cite{Kovacik:2016dwp,Kovacik:2017sbw,Kovacik:2018vvx}, and in connections between non-commutative quantum mechanics and anti-hermitian 
$\mathbb{H}$QM as well \cite{Soloviev:2016qsx,Soloviev:2017nwk,Gracia-Bondia:2017fai,Carinena:2009ug}.

 In summary, the theoretical novelties presented in this article are the expression of electromagnetic fields in terms of quaternionic vectors, the direct link between the classical results and the quantum theory, the existence of a self-interaction due to the vector potential, and the existence of magnetic  monopole solutions free of singularities.

For later use, we define the vector product between the quaternionic vectors $\,\bm{\mathcal{X}}=\bm X_0+\bm X_1j\,$ and $\,\bm{\mathcal{Y}}=\bm Y_0+\bm Y_1j\,$ to be
\begin{equation}\label{i02}
\bm{\mathcal{X}\times\mathcal{Y}}=\bm{X_0\times Y_0}-\bm{X_1\times \overline Y_1}+\big(\bm{X_0\times Y_1}+\bm{X_1\times\overline Y_0}\big)j,
\end{equation}
and consequently $\,\bm{\mathcal{Y}\times\mathcal{X}}\neq -\bm{\mathcal{X}\times\mathcal{Y}}.\,$ Other useful relation is
\begin{equation}\label{i03}
\overline{ \bm{\mathcal{X}\times\mathcal{Y}}}\,=\,-\,\bm{\overline{\mathcal{Y}}\times\overline{\mathcal{X}}}.
\end{equation}
In order to simplify the notation, we define the symmetrized and anti-simmetrized products
\begin{equation}\label{i04}
\big( X,\, Y\big)=\frac{1}{2}\Big(X\star \overline Y+  Y\star \overline X\Big)\qquad\mbox{and}\qquad
 \big[X,\, Y\big]=\frac{1}{2}\Big( X\star \overline Y- Y\star \overline X\Big),
\end{equation}
where three different products involving scalars and vectors are considered
\begin{equation}\label{i05}
 X\star Y\in\Big\{\mathscr{X}\mathscr{Y},\;\mathscr{X}\bm{\mathcal{Y}},\;\bm{\mathcal{X}\cdot\mathcal{Y}} \Big\}.
\end{equation}
Of course, the symmetrized products are real and the anti-symmetrized products are pure imaginary.
On the other hand, in the case of the vector product we have that
\begin{equation}\label{i06}
 \big\{\bm{\mathcal{X},\,\mathcal{Y}}\big\}_-=\frac{1}{2}\big(\bm{\mathcal{X}\times\overline{\mathcal{Y}}}-\bm{\mathcal{Y}}\times\bm{\overline{\mathcal{X}}}\big),
\qquad\qquad
\big\{\bm{\mathcal{X},\,\mathcal{Y}}\big\}_+=\frac{1}{2}\big(\bm{\mathcal{X}\times\overline{\mathcal{Y}}}+\bm{\mathcal{Y}\times\overline{\mathcal{Y}}}\big),
\end{equation}
and (\ref{i03}) imposes that $\,\big\{\bm{\mathcal{X},\,\mathcal{Y}}\big\}_-\,$  is real and that $\,\big\{\bm{\mathcal{X},\,\mathcal{Y}}\big\}_+\,$ is pure imaginary.

\section{\sc Quaternionic electrodynamics\label{M}}
Admitting the existence of magnetic monopoles, the quaternionic electromagnetic field equations are such that
\begin{align}\nonumber
&\bm{\nabla\cdot \mathcal{E}}=4\pi\,\mathscr{P}\qquad\qquad\qquad\qquad\qquad\;\;\;\;\bm{\nabla\cdot \mathcal{B}}=4\pi\,\mathscr{P}_M\\
\label{m2}
&\bm{\nabla\times \mathcal{E}}=-\frac{4\pi}{c}\,\bm{\mathcal{J}}_M-\frac{1}{c}\frac{\partial \bm{\mathcal{B}}}{\partial t}\qquad\qquad\;\;\;\bm{\nabla\times\mathcal{B}}=\frac{4\pi}{c}\,\bm{\mathcal{J}}+\frac{1}{c}\frac{\partial \bm{\mathcal{E}}}{\partial t},
\end{align}
where the electric and magnetic fields $\bm{\mathcal{E}}$ and $\,\bm{\mathcal{B}}$, the electric and magnetic charge densities $\mathscr{P}$ and $\mathscr{P}_M$,  and the electric and magnetic charge currents 
$\bm{\mathcal{J}}$ and $\bm{\mathcal{J}}_M$ are all pure quaternionic. The above equations are invariant to quaternionic conjugation and to the duality symmetry
\begin{equation}\label{m01}
\bm{\mathcal{E}}\to\bm{\mathcal{B}},\qquad\mathscr{P}\to\mathscr{P}_M,\qquad\bm{\mathcal{J}}\to\bm{\mathcal{J}}_M,\qquad\qquad
\bm{\mathcal{B}}\to-\bm{\mathcal{E}},\qquad\mathscr{P}_M\to-\mathscr{P},\qquad\bm{\mathcal{J}}_M\to-\bm{\mathcal{J}},
\end{equation}
and also to the symmetry operation
\begin{align}\nonumber
&\bm{\mathcal{E}}\to\bm{\mathcal{E}}+\bm{\mathcal{B}},\qquad\mathscr{P}\to\mathscr{P}+\mathscr{P}_M,
\qquad\;\;\;\;\bm{\mathcal{J}}\to\bm{\mathcal{J}}+\bm{\mathcal{J}}_M,\\ 
\label{m02}
&\bm{\mathcal{B}}\to\bm{\mathcal{B}}-\bm{\mathcal{E}},\qquad\mathscr{P}_M\to\mathscr{P}_M-\mathscr{P},\qquad
\bm{\mathcal{J}}_M\to\bm{\mathcal{J}}_M-\bm{\mathcal{J}}.
\end{align}
Applying the divergence theorem to the magnetic field, we obtain
\begin{equation}\label{m3}
\oint\bm{\mathcal{B}\cdot}d\bm{\sigma}\;=\;\oint \bm{\mathcal{A}\cdot}d\bm\ell-\oint \bm{\mathcal A\times\mathcal A\cdot}d\bm\sigma\,=\;4\pi\,\mathscr{Q}_M,
\end{equation}
where $d\bm\sigma$ is the surface element over the normal direction of the surface and $d\bm\ell$ is the length element of the closed 
path around a singularity of $\bm{\mathcal{A}}$. It has been also used that
\begin{equation}\label{m4}
 \mathscr{Q}_M=\int_V\mathscr{P}_M dv,
\end{equation}
where $\mathcal{Q}_M$ is the pure quaternionic magnetic charge obtained after integration over the volume of the whole space $V$. 
Equations (\ref{m3}-\ref{m4}) show that the quaternionic magnetic field may admit the existence of magnetic charges for non-singular vector potentials, an important difference to the current formulation 
of magnetic monopoles, where the vector potential is necessarily singular and the set of singularities
comprises the Dirac string. 
\subsection{\sc Gauge transformations\label{GT}}
Equation  (\ref{i2}) and the divergence of the magnetic field (\ref{m2})  give
\begin{equation}\label{gt01}
\bm{\nabla\cdot\big(\mathcal{A}\times\mathcal{A}\big)}=-\,4\pi\,\mathscr{P}_M,
\end{equation}
and thus we observe the dynamical origin of the magnetic charge density. This hypothesis is reinforced after calculating the curl of the electric field, namely
\begin{equation}\label{gt02}
\bm{\nabla\times}\left(\bm{\mathcal{E}}+\frac{1}{c}\frac{\partial\bm A}{\partial t}\right)=-\frac{4\pi}{c}\bm{\mathcal{J}}_M+\frac{1}{c}\frac{\partial}{\partial t}\bm{\mathcal{A}\times\mathcal A}.
\end{equation}
Imposing the right hand side of (\ref{gt02}) to zero, we obtain
\begin{equation}\label{gt03}
\bm{\mathcal{J}_M}=\frac{1}{4\pi}\frac{\partial}{\partial t}\bm{\mathcal{A}\times\mathcal A},
\end{equation}  
and the magnetic current is dynamically also generated by the self-interaction of $\bm{\mathcal{A}}$. Finally, equation (\ref{gt01}) and the divergence of (\ref{gt03})  produce the continuity equation
\begin{equation}
\bm{\nabla\cdot\mathcal{J}_M}+\frac{\partial\mathscr{P}_M}{\partial t}=0.
\end{equation}
Consequently, (\ref{gt02}) implies the usual electric field expression
\begin{equation}\label{gt04}
\bm{\mathcal{E}}=-\bm\nabla\Phi-\frac{1}{c}\frac{\partial\bm{\mathcal{A}}}{\partial t},
\end{equation}
where $\Phi$ is a pure imaginary quaternionic scalar potential. Finally, the curl of the magnetic field gives
\begin{equation}\label{gt05}
\nabla^2\bm{\mathcal{A}}-\frac{1}{c^2}\frac{\partial^2\bm{\mathcal{A}}}{\partial t^2}=-\frac{4\pi}{c}\bm{\mathcal{J}}-
\bm{\nabla\times\big(\mathcal{A}\times\mathcal{A}\big)},
\end{equation}
where we imposed the Lorentz gauge condition 
\begin{equation}\label{gt06}
\frac{1}{c}\frac{\partial\Phi}{\partial t}+\bm{\nabla\cdot\mathcal{A}}=0.
\end{equation}
The interaction term in the right hand side of (\ref{gt05}) is a physical novelty that is absent in the usual electromagnetic theory as well as
in the Dirac monopole theory, and it is an interesting theme for future research.
The gauge transformations in this quatenionic theory are then
\begin{equation}\label{gt07}
\bm{\mathcal{A}}\to\bm{\mathcal{A}}^\prime+\bm\nabla\Psi\qquad\mbox{and}\qquad\Phi\to\Phi^\prime-\frac{1}{c}\frac{\partial\Psi}{\partial t}
\end{equation} 
where $\Psi$ is a quaternionic function constrained according to
\begin{equation}\label{gt08}
\bm{\mathcal{A}^\prime\times\nabla}\Psi+\bm\nabla\Psi\times\bm{\mathcal{A}^\prime}+\bm\nabla\Psi\times\bm\nabla\Psi=0.
\end{equation}
 We observe that (\ref{gt04}) is different from the usual expression of generalized Dirac Maxwell equations \cite{Majernik:1999rad,Weng:2015qoa,Hong:2019azp,Chanyal:2017sjj,Kansu:2020atu}, where both of the fields have scalar and vector potentials. In the quaternionic formulation contained in this article, the magnetic field 
(\ref{i2}) is essentially different from the electric field because of the self-interaction term, and consequently
(\ref{gt04}) holds in accordance to Maxwell electrodynamics. The possible existence of dyons in this quaternionic theory is another exciting  question to be responded in future research.
 Our next interest is to analyze the
dynamics of the electromagnetic force. Surprisingly, there are two possibilities in agreement to our assumptions. Let us consider the first of them.

\subsection{\sc The real dynamics}
The non-commutativity of quaternions sets an ordering ambiguity when the formulation of physical laws is tried.   We propose an electromagnetic force law of symmetrized quaternionic products, so that
\begin{equation}\label{m013}
\bm{\mathcal{F}}=-\frac{1}{2}\left[\mathscr{P}\bm{\mathcal{E}}+\bm{\mathcal{E}}\mathscr{P}+\mathscr{P}_M\bm{\mathcal{B}}+\bm{\mathcal{B}}\mathscr{P}_M+
\frac{1}{c}\Big(\bm{\mathcal{J}}\times\bm{\mathcal{B}}\,-\,\bm{\mathcal{B}}\times\bm{\mathcal{J}}\;-\;
\bm{\mathcal{J}}_M\times\bm{\mathcal{E}}\,+\,\bm{\mathcal{E}}\times\bm{\mathcal{J}}_M\Big)\right],
\end{equation}
where the negative sign is justified for the fields are pure imaginary. The force can be rewritten using the definitions (\ref{i04}-\ref{i06}) as
\begin{equation}\label{m13}
\bm{\mathcal{F}}\,=\,\big(\mathscr{P},\,\bm{\mathcal{E}}\big)\,+\,\big(\mathscr{P}_M,\,\bm{\mathcal{B}}\big)\,+\,
\frac{1}{c}\left(\;\big\{\bm{\mathcal{J}},\,\bm{\mathcal{B}}\big\}_-\;-\;\big\{\bm{\mathcal{J}}_M\,\,\bm{\mathcal{E}}\big\}_-\right).
\end{equation}
More importantly, (\ref{m13}) is real and consequently the quaternionic fields and charges generate a real dynamics
where the force is the only physical observable, while the fields are uniquely theoretical devices.

The electrodynamic force permits us to ascertain the energy conservation. The total work $\,\mathcal W\,$ over a flux of charged particles that performs a real velocity $\bm u\,$ must satisfy
\begin{equation}\label{m6}
\frac{d\mathcal W}{dt}=\int \bm{\mathcal{F}\cdot u}\,dv=-\frac{1}{2}\int\Big(\,\bm{\mathcal{E}\cdot\mathcal{ J}}+\bm{\mathcal{J}\cdot\mathcal{ E}}\;+\; \bm{\mathcal{B}\cdot\mathcal{J}_M}+\bm{\mathcal{J}_M\cdot\mathcal{B}}\,\Big)dv,
\end{equation}
where $\bm{\mathcal F}$ is the Lorentz force over a single particle and the integral is performed over the whole space. 
Obtaining $\,\bm{\mathcal{J}}\,$ and $\,\bm{\mathcal{J}_M}\,$ from the field equations (\ref{m2}) and using the quaternionic vector identity
\begin{equation}\label{m8}
\bm{\nabla\cdot\big(\mathcal{E}\times\mathcal{B}\big)}=\bm{\big(\nabla\times\mathcal{E}\big)\cdot\mathcal{B}}\,-\,\bm{\mathcal{E}\cdot\big(\nabla\times\mathcal{B}\big)},
\end{equation}
we obtain the conservation law 
\begin{equation}
\frac{\partial \mathcal{U}}{\partial t}+\bm{\nabla\cdot\mathcal{S}}= 
\,-\,\frac{1}{2}\Big(\bm{\mathcal{E}\cdot\mathcal{J}}+\bm{\mathcal{J}\cdot\mathcal{E}}\Big)\,-\,
\frac{1}{2}\Big(\bm{\mathcal{B}\cdot\mathcal{J}_M}+\bm{\mathcal{J}}_M\bm{\cdot\mathcal{B}}\Big),
\end{equation}
which, according to (\ref{i04}-\ref{i06}) is
\begin{equation}\label{m9}
\frac{\partial \mathcal{U}}{\partial t}+\bm{\nabla\cdot\mathcal{S}}
\,=\,\big(\bm{\mathcal{E},\,\mathcal{J}}\big)\,+\,\big(\bm{\mathcal{B},\,\mathcal{J}_M}\big).
\end{equation}
Equation (\ref{m9}) is the quaternionic version of the Poynting theorem, where the Poynting vector $\bm{\mathcal{S}}$ and the total
energy density $\,\mathcal U\,$ are
\begin{equation}\label{m10}
\bm{\mathcal{S}}=c\,\frac{\bm{\mathcal{E}\times\mathcal{B}}-\bm{\mathcal{B}\times\mathcal{E}}}{8\pi}\,=\,\frac{c}{4\pi}\big\{\bm{\mathcal{B},\,\mathcal{E}}\big\}_-
\qquad\mbox{and}\qquad \mathcal{U}=\frac{|\bm{\mathcal{E}}|^2+|\bm{\mathcal{B}}|^2}{8\pi}.
\end{equation}
The energy flux vector $\,\bm{\mathcal{S}},\,$ the energy density $\,\mathcal{U},\,$ and equation (\ref{m9}) are real and can
be interpreted  in the same spirit of classical electrodynamics, which is immediately recovered by setting to zero the pure quaternionic components, and   indicates that the quaternionic result consistently generalizes the real dynamics. The 
energy flux $\,\bm{\mathcal{S}}\,$ is expected to be orthogonal to $\,\bm{\mathcal{E}}\,$ and to $\,\bm{\mathcal{B}}.\,$ Using the scalar product (\ref{i04}), we obtain
\begin{equation}
\big(\bm{\mathcal{S}},\,\bm{\mathcal{E}}\big)=\big(\bm{\mathcal{S}},\,\bm{\mathcal{B}}\big)=0
\end{equation}
what reinforces that the physical content of the theory is real valued. 
We can further advance considering the total momentum of the particles after integrating (\ref{m13}) over the whole space
\begin{equation}\label{m14}
 \frac{d\bm{\mathcal{P}}_P}{dt}=\int \bm{\mathcal{F}}dv.
\end{equation}
Thus, using the field equations (\ref{m2}) to eliminate the charges and currents, we obtain
\begin{align}\nonumber
\frac{d\bm{\mathcal{P}}_P}{dt}\,=\,&-\frac{1}{8\pi c}\int\frac{\partial}{\partial t}\Big(\bm{\mathcal{B}\times\mathcal{E}}-\bm{\mathcal{E}\times\mathcal{B}}\Big)dv\,-\,
\frac{1}{8\pi}\int\Big[\big(\bm{\nabla\cdot\mathcal{E}}\big)\bm{\mathcal{E}}\,+\,\bm{\mathcal{E}}\big(\bm{\nabla\cdot\mathcal{E}}\big)\,+\,\big(\bm{\nabla\cdot\mathcal{B}}\big)\bm{\mathcal{B}}\,+\,\bm{\mathcal{B}}\big(\bm{\nabla\cdot\mathcal{B}}\big)+
\\
\label{m15}
&+\big(\bm{\nabla\times\mathcal{E}}\big)\times\bm{\mathcal{E}}\,-\,\bm{\mathcal{E}\times}\big(\bm{\nabla\times\mathcal{E}}\big)\,+\big(\bm{\nabla\times\mathcal{B}}\big)\bm{\times\mathcal{B}}\,-\,\bm{\mathcal{B}\times}\big(\bm{\nabla\times\mathcal{B}}\big)\Big]dv.
\end{align}
In order to simplify (\ref{m15}), we use the Leibnitz rule $\,\bm{\nabla\cdot}\big(X\bm{Y}\big)=(\bm\nabla X)\bm{\cdot Y}+X(\bm{\nabla\cdot Y})\,$  
for a quaternionic scalar $\,X\,$ and a quaternionic vector $\,\bm Y\,.$ Thus,
\begin{equation}\label{m151}
\bm{\nabla\cdot}\big(\mathcal{E}_i\bm{\mathcal{E}}+\bm{\mathcal{E}}\mathcal{E}_i\big)=\bm{\mathcal{E}\cdot}\bm\nabla\mathcal{E}_i+\bm\nabla\mathcal{E}_i\bm{\cdot\mathcal{E}}+(\bm{\nabla\cdot\mathcal{E}})\mathcal{E}_i+\mathcal{E}_i\bm{\nabla\cdot\mathcal{E}}.
\end{equation}
Additionally, using
\begin{equation}\label{m16}
\Big(\bm\nabla(\bm{X\cdot Y})\Big)_i=\Big(\bm{X\cdot\nabla}\Big)Y_i+\Big(\bm\nabla X_i\Big)\bm{\cdot Y}+
\Big(\bm{X\times(\nabla\times Y)}\Big)_i-\Big((\bm{\nabla\times X})\bm{\times Y}\Big)_i
\end{equation}
with $X=Y=\mathcal{E}$, we obtain
\begin{equation}\label{m161}
\bm{\nabla\cdot}\Big(\mathcal{E}_i\bm{\mathcal{E}}+\bm{\mathcal{E}}\mathcal{E}_i\Big)-\Big(\bm{\nabla}\mathcal{E}^2\Big)_i=
\big(\bm{\nabla\cdot\mathcal{E}}\big)\mathcal{E}_i\,+\,\mathcal{E}_i\big(\bm{\nabla\cdot\mathcal{E}}\big)
+\Big(\big(\bm{\nabla\times\mathcal{E}}\big)\times\bm{\mathcal{E}}\Big)_i\,-\,\Big(\bm{\mathcal{E}\times}\big(\bm{\nabla\times\mathcal{E}}\big)\Big)_i.
\end{equation}
After integrating over the whole space,  the divergence theorem over the three directions produces
\begin{equation}
\int\left[  \bm{\nabla\cdot}\Big(\mathcal{E}_i\bm{\mathcal{E}}+\bm{\mathcal{E}}\mathcal{E}_i\Big)-\Big(\bm{\nabla}\mathcal{E}^2\Big)_i\,\right] dv\;\to\;
\oint \Big[\big(\bm{\mathcal{E}\cdot n}\big)\bm{\mathcal{E}}+\bm{\mathcal{E}}\big(\bm{\mathcal{E}\cdot n}\big)-\mathcal{E}^2\bm n\Big]da,
\end{equation}
where $\bm n$ is the unitary vector that is normal to the integration surface, where $d\bm a=\bm n da$.
An identical calculation can be done for $\bm{\mathcal{B}}$. All these results and (\ref{m151}) give
\begin{equation}\label{m17}
 \frac{d}{dt}\big(\bm{\mathcal{P}}_P+\bm{\mathcal{P}}_F\big)_k\,=\,-\sum_\ell\oint\mathcal T_{k\ell}n_\ell da
\end{equation}
where $\bm{\mathcal{P}}_F$ is the momentum stored in the fields and $\mathcal{T}_{k\ell}$ is the quaternionic
stress-energy tensor, namely
\begin{equation}\label{m18}
 \bm{\mathcal{P}}_F=\frac{1}{8\pi c}\Big(\bm{\mathcal{B}\times\mathcal{E}}-\bm{\mathcal{E}\times\mathcal{B}}\Big)=-\frac{1}{c^2}\bm{\mathcal{S}}\qquad\mbox{and}\qquad 
 \mathcal T_{k\ell}=\frac{1}{4\pi }\Bigg[\,\frac{|\mathcal E|^2+|\mathcal B|^2}{2}\delta_{k\ell}\,-\,\big(\mathcal{E}_{k},\,\mathcal{E}_{\ell}\big)\,-\,\big(\mathcal{B}_{k},\,\mathcal{B}_{\ell}\big)\,\Bigg].
\end{equation}
All the results strikingly agree to the usual real results. However, with additional degrees of freedom, which describe a novel classical physics, and may be needful for quantum phenomena, 
particularly in the case of $\mathbb{H}$QM. In the sequel we describe an entirely different theory, in terms of imaginary degrees of freedom.

\subsection{\sc The imaginary dynamics\label{tid}}
The symmetry transformations (\ref{m01}-\ref{m02}) can guide us towards a pure imaginary electromagnetic force law
\begin{equation}\label{m19}
\bm{G}=\frac{1}{2}\left[\mathscr{P}_M\bm{\mathcal{E}}-\bm{\mathcal{E}}\mathscr{P}_M-\mathscr{P}\bm{\mathcal{B}}+\bm{\mathcal{B}}\mathscr{P}
+\frac{1}{c}\Big(\bm{\mathcal{J}}\times\bm{\mathcal{E}}\,+\,\bm{\mathcal{E}}\times\bm{\mathcal{J}}\;+\;
\bm{\mathcal{J}}_M\times\bm{\mathcal{B}}\,+\,\bm{\mathcal{B}}\times\bm{\mathcal{J}}_M\Big)\right],
\end{equation}
which can be rewritten as
\begin{equation}\label{m20}
\bm{G}=\big[\mathscr{P},\bm{\mathcal{B}}\big]-\big[\mathscr{P}_M,\,\bm{\mathcal{E}}\big]
-\frac{1}{c}\Big(\big\{\bm{\mathcal{J}},\,\bm{\mathcal{E}}\big\}_+\,+\,\big\{\bm{\mathcal{J_M}},\,\bm{\mathcal{B}}\big\}_+\Big).
\end{equation}
The conservation law will be
\begin{equation}\label{m11}
\frac{\partial \mathcal{Y}}{\partial t}+\bm{\nabla\cdot\mathcal{C}}= 
\,\,\frac{1}{2}\Big(\bm{\mathcal{E}\cdot\mathcal{J}_M}-\bm{\mathcal{J}_M\cdot\mathcal{E}}\Big)\,+\,
\frac{1}{2}\Big(\bm{\mathcal{J}}\bm{\cdot\mathcal{B}-\bm{\mathcal{B}\cdot\mathcal{J}}}\Big)
\end{equation}
and using (\ref{i04}-\ref{i06}) we get
\begin{equation}\label{m210}
\frac{\partial \mathcal{Y}}{\partial t}+\bm{\nabla\cdot\mathcal{C}}= 
\big[\bm{\mathcal{B}},\,\bm{\mathcal{J}}\big]\,-\,\big[\bm{\mathcal{E},\,\mathcal{J}_M}\big],
\end{equation}
where the momentum flux vector $\bm{\mathcal{C}}$ and the energy density $\,\mathcal{Y}\,$ are such that
\begin{equation}\label{m12}
\bm{\mathcal{C}}=c\,\frac{\bm{\mathcal{E}\times\mathcal{E}}+\bm{\mathcal{B}\times\mathcal{B}}}{8\pi}
\qquad\mbox{and}\qquad \mathcal{Y}=\frac{\bm{\mathcal{B}\cdot\mathcal{E}}-\bm{\mathcal{E}\cdot\mathcal{B}}}{8\pi}\,=\,\frac{1}{4\pi}\big[\bm{\mathcal{E},\,\mathcal{B}}\big].
\end{equation}
Hence the energy flux $\,\bm{\mathcal{C}}\,$ is defined by the self-interaction between the quaternionic fields, and the total energy $\mathcal{Y}\,$ is obtained from the interaction between
the fields. A totally different dynamical system, when compared to the previous real dynamics. If we define a anti-commutative scalar product, the energy flux
$\bm{\mathcal{C}}$ is self-orthogonal, and if a commutative scalar product is defined following (\ref{i04}), there not a natural candidate for the orthogonal direction
to the energy flux, and therefore we postpone this point to future research.
Despite of this, the conservation of momentum can also be obtained in the same fashion of the real dynamics by using the charges and currents from the field equations (\ref{m2}) and the quaternionic vector identities 
(\ref{m16}-\ref{m161}) such that
\begin{equation}
G_i\,=\,\frac{1}{8\pi}\left\{-\frac{1}{c}\frac{\partial}{\partial t}\Big(\bm{\mathcal{E}\times\mathcal{E}}+\bm{\mathcal{B}\times\mathcal{B}}\Big)_i+\bm{\nabla\cdot}\Big(\mathcal{E}_i\bm{\mathcal{B}}
-\bm{\mathcal{B}}\mathcal{E}_i-\mathcal{B}_i\bm{\mathcal{E}}+\bm{\mathcal{E}}\mathcal{B}_i\Big)+\Big[\bm\nabla\big(\bm{\mathcal{E}\cdot\mathcal{B}}-\bm{\mathcal{B}\cdot\mathcal{E}}\big)\Big]_i
\right\}.
\end{equation}
Integrating $\bm G$ in the whole space, such as in (\ref{m14}), we obtain the time devivative of the particle momentum $\,\bm{\mathcal{K}}_P\,$   and consequently the conservation law
\begin{equation}\label{m200}
 \frac{d}{dt}\big(\bm{\mathcal{Q}}_P+\bm{\mathcal{Q}}_F\big)_k\,=\,\sum_\ell\oint\mathcal D_{k\ell}n_k da.
\end{equation}
$\bm{\mathcal{Q}}_F$ is the momentum stored in the fields and $\mathcal{D}_{k\ell}$ is the quaternionic
stress-energy tensor, namely
\begin{equation}\label{m21}
 \bm{\mathcal{Q}}_F=\frac{1}{8\pi c}\Big(\bm{\mathcal{E}\times\mathcal{E}}+\bm{\mathcal{B}\times\mathcal{B}}\Big)\,=\,\frac{1}{c^2}\bm{\mathcal{C}}\qquad\mbox{and}\qquad 
 \mathcal D_{k\ell}=\frac{1}{4\pi }\Big(\big[\mathcal{B}_{k},\,\mathcal{E}_{\ell}\big]+\big[\mathcal{B}_{\ell},\,\mathcal{E}_{k}\big]+
\big[\bm{\mathcal{B},\,\mathcal{E}}\big]\delta_{k\ell}\Big).
\end{equation}
The imaginary dynamics is something new and fascinating. We can speculate many hypothesis about it, such as the description of either anti-particles or quantum effects, and all of these are interesting directions for future research.

\section{\sc Magnetic monopole solutions}
The constructed quaternionic electrodynamics ($\mathbb{H}$ED) of the previous section generalizes the real theory, and
we can investigate every aspect of the theory without the fear of inconsistency. 
We choose several examples of magnetic monopoles in accordance to (\ref{m2}) as the features to be explored in the following subsections.
\subsection{\sc radial monopole field\label{R}}
The field of a monopole is expected to be
\begin{equation}\label{r1}
\bm{\mathcal{B}}\propto \frac{\bm{\hat r}}{r^2},
\end{equation}
where $\,r\,$ is the spherical coordinate radius. In order to accomplish such a magnetic field, we propose the quaternionic vector potential $\,\bm{\mathcal{A}}\,$ to be comprised of
\begin{equation}\label{r2}
 \bm\alpha=\bm 0\qquad\mbox{and}\qquad\bm\beta=\,\frac{1}{r}\left(g_\theta\frac{\sin\theta}{1+\cos\theta}\,\bm{\hat\theta}\,-\,g_\phi\frac{1+\cos\theta}{\sin\theta}\bm{\hat{\phi}}\right),
\end{equation}
where $\,\theta\,$ and $\,\phi\,$ are respectively the polar and azimuthal spherical coordinates and where $\,g_\theta\,$ and $\,g_\phi\,$ are complex magnetic charges. From (\ref{r2}),
\begin{equation}\label{r3}
\bm{\mathcal{B}}\,=\,\bm{\bar\beta\times\beta}\,+\,\bm{\nabla\times\beta}j\,=\,\Big(g_\phi\bar g_\theta-g_\theta\bar g_\phi\,+\,g_\phi j\Big)\frac{\bm{\hat r}}{r^2}.
\end{equation}
And from from (\ref{m3}) the magnetic charge is
\begin{equation}\label{r4}
\oint\bm{\beta\cdot }d\bm\ell+\int\bm{\bar\beta\times\beta\cdot}d\bm\sigma=4\pi\Big(g_\phi +g_\phi\bar g_\theta-g_\theta\bar g_\phi\Big),
\end{equation}
where the Stokes theorem has been used. This example comprises the simplest quaternionic magnetic monopole one can imagine, and it is in closed
relation to the Dirac monopole, which explicitly appears in the singularities at $\theta=0$  and $\theta=\pi/2$, consequently characterizing  Dirac strings.
Let us now entertain two variations of this monopole solution.

\paragraph{\underline{ Magnetic monopole inside an electrically charged rotating spherical shell}}
A spherical shell of radius $R$ spins at angular velocity $\,\bm\omega=\omega\bm{\hat z}\,$ and carries a uniformly distributed electrical charge of surface density $\mathscr{S}$ and
a magnetic monopole identical to that obtained in the previous example lays at the center of the sphere. The charged  spherical surface generates a magnetic field whose 
vector potential is accordingly
\begin{equation}\label{r5}
\bm\alpha=\left\{
\begin{array}{ll}
\alpha_0\sin\theta\frac{r}{R}\bm{\hat\phi} &\qquad\mbox{if}\qquad r\leq R\\
\alpha_0\sin\theta\left(\frac{R}{r}\right)^2\bm{\hat\phi} &\qquad\mbox{if}\qquad r\geq R,
\end{array}
\right.
\qquad\mbox{where}\qquad \alpha_0=\frac{(4\pi R)^2\omega\mathscr{S}}{3c}
\end{equation}
and the magnetic field is
\begin{equation}\label{r6}
\bm B=\bm{\nabla\times\alpha}=
\left\{
\begin{array}{ll}
\frac{2\alpha_0}{R}\big(\cos\theta\,\hat{\bm r}-\sin\theta\,\hat{\bm\theta}\big) &\qquad\mbox{if}\qquad r\leq R\\&\\
\frac{\alpha_0 R^2}{r^3}\big(2\cos\theta\,\hat{\bm r}+\sin\theta\,\hat{\bm\theta}\big) &\qquad\mbox{if}\qquad r\geq R.
\end{array}
\right.
\end{equation}
We notice that the magnetic field is constant inside the spherical shell because $\bm{\hat z}=\cos\theta\,\bm{\hat r}-\sin\theta\,\hat{\bm\theta}.\,$ 
Thus, taking $\bm\beta$ from (\ref{r2}) the interaction term is simply
\begin{equation}\label{r7}
\bm{\alpha\times\beta}=\left\{
\begin{array}{ll}
-\frac{\alpha_0}{R}\big(1-\cos\theta\big)g_\theta \,\bm{\hat r} &\qquad\mbox{if}\qquad r\leq R\\&\\
-\alpha_0 R^2\big(1-\cos\theta\big)\frac{g_\theta}{r^3}\bm{\hat r} &\qquad\mbox{if}\qquad r\geq R.
\end{array}
\right.
\end{equation}
This interaction is regular on the whole space term contributes and it decreases
at a rate identical to real magnetic field outside the spherical shell. On the other hand, the charge associated to the interaction is
\begin{equation}\label{r8}
\int\bm{\alpha\times\beta}\bm\cdot d\bm\sigma=\left\{
\begin{array}{ll}
-\frac{4\pi\alpha_0}{R}g_\theta r^2 &\qquad\mbox{if}\qquad r\leq R\\&\\
-4\pi\alpha_0 R^2\frac{g_\theta}{r} &\qquad\mbox{if}\qquad r\geq R.
\end{array}
\right.
\end{equation}
We observe that the charge of the interaction is a non-constant function of $r$. This fact seems reasonable because there is no physical
charge associated to the interaction, it is in fact a virtual charge that depends on $\bm\alpha$ and $\bm\beta$, and it may be controlled by
an external parameter, such as the angular velocity, the surface density, or accordingly the distance from the monopole. We can interpret that the
interaction between the fields described in terms of the interaction of the potential polarizes the space and generate dynamical magnetic charges.
\paragraph{\underline{ Magnetic monopole inside an infinite cylindrical solenoid}}

The magnetic field is constant inside an infinite solenoid of radius $R$ where flows a constant electric current. In cylindrical coordinates, we have
\begin{equation}\label{r9}
\bm B=\left\{
\begin{array}{ll}
B_0\bm{\hat z} &\qquad\mbox{if}\qquad \rho\leq R\\
0 &\qquad\mbox{if}\qquad \rho\geq R,
\end{array}
\right.
\qquad\mbox{so that}\qquad B_0=\frac{\Phi_0}{\pi R^2},
\end{equation}
where $\Phi_0$ is the total magnetic flux inside the solenoid and $\rho\,$ is the cylindrically symmetric radius. The magnetic vector potential for this system is
\begin{equation}\label{r10}
\bm\alpha=\left\{
\begin{array}{ll}
\frac{B_0}{2}\,\rho\,\bm{\hat\phi}&\qquad\mbox{if}\qquad \rho\leq R\\&\\
\frac{B_0 R^2}{2}\,\frac{1}{\rho}\,\bm{\hat\phi} &\qquad\mbox{if}\qquad \rho\geq R.
\end{array}
\right.
\end{equation}
Let us introduce a magnetic monopole inside this solenoid. The azimuthal angle is common to the spherical coordinate, and using $\rho=r\sin\theta$ we immediately obtain 
the regular contribution to the quaternionic magnetic field
\begin{equation}\label{r11}
\bm{\alpha\times\beta}=\left\{
\begin{array}{ll}
-\frac{B_0}{2}g_\theta\big(1-\cos\theta\big)\,\,\bm{\hat r} &\qquad\mbox{if}\qquad r\sin\theta\leq R\\&\\
-\frac{B_0}{2}\frac{g_\theta}{1+\cos\theta}\left(\frac{R}{r}\right)^2\bm{\hat r} &\qquad\mbox{if}\qquad r\sin\theta\geq R,
\end{array}
\right.
\end{equation}
and observe that the field is regular outside the solenoid, although $\theta\to\pi/2$ may be reached within the limit of infinite $r$. On the other hand, 
\begin{equation}\label{r12}
\int\bm{\alpha\times\beta}\bm\cdot d\bm\sigma=\left\{
\begin{array}{ll}
-2 \Phi_0g_\theta \,\left(\frac{r}{R}\right)^2 &\qquad\mbox{if}\qquad r\sin\theta\leq R\\&\\
-\Phi_0 g_\theta\left[2-\displaystyle{\lim_{\cos\theta\to -1}}\ln\big(1+\cos\theta\big)\right] &\qquad\mbox{if}\qquad r\sin\theta\geq R.
\end{array}
\right.
\end{equation}
The interpretation is identical to the former case, where the virtual charge depends on parameters like $\Phi_0$ and $r$. Although the magnetic field is zero 
outside the solenoid, the vector potential induces the existence of a charge in this region, and consequently there is a field given by (\ref{r11}). Another
novel result that cannot be obtained in the real electrodynamic theory.

\subsection{\sc cylindric monopole field\label{C}}
We expect a field of an infinite magnetically charged wire behaves like the electrically charged wire. Consequently,
\begin{equation}\label{c1}
\bm{\mathcal{B}}\propto \frac{\bm{\hat \rho}}{\rho},
\end{equation}
where $\rho$ is the cylindrically symmetric radial coordinate. The simplest possibility to the components of the quaternionic vector potential seem to be
\begin{equation}\label{c2}
 \bm\alpha=\bm 0\qquad\mbox{and}\qquad\bm\beta=\,\sqrt{2}\left(\frac{\lambda_\phi}{\rho}\bm{\hat\phi}\,+\,\lambda_z\bm{\hat z}\right),
\end{equation}
where $\lambda_\theta$ and $\lambda_\phi$ are linear charge densities along the wire. Thus,
\begin{equation}\label{c3}
 \bm{\nabla\times\beta}=\bm 0\qquad\mbox{and}\qquad\bm{\bar\beta\times\beta}=2\frac{\lambda_z\bar\lambda_\phi-\lambda_\phi\bar\lambda_z}{\rho}\bm{\hat \rho}.
\end{equation}
We stress that $\,\int\bm{\nabla\times\beta\cdot}d\sigma=\int\bm{\beta\cdot}d\bm\ell=0\,$ because the contributions from both edges of the cylindrical surface cancel out. Consequently, 
integrating over a cylindrial surface of radius $\rho$ and axis along $\,z\in(-L/2,\,L/2)$, we have
\begin{equation}\label{c4}
 \bm{\mathcal B}=\frac{g_\phi\bar g_z-g_z\bar g_\phi}{\rho}\bm{\hat \rho}
\qquad\mbox{and}\qquad
\int\bm{\bar\beta\times\beta\cdot}d\bm\sigma=4\pi\Big(g_z\bar g_\phi-g_\phi \bar g_z\Big),
\end{equation}
where $g_\theta=\lambda_\theta L$ and $g_\phi=\lambda_\phi L$ are complex charges.  The field of the magnetically charged wire is identical to the field of the electrically charged wire. However, there are no Dirac strings
connecting the monopoles to infinity. This is a novel result that is not observed in the usual electromagnetic theory.
\paragraph{\underline{ Magnetically charged wire inside an infinite cylindrical solenoid}}
The magnetic vector potential $\bm\alpha$ from (\ref{r9}-\ref{r10}) and the vector potential $\bm\beta$ from (\ref{c2}) enable us to obtain
\begin{equation}\label{c5}
\bm{\alpha\times\beta}=\left\{
\begin{array}{ll}
\frac{B_0}{\sqrt{2}}\lambda_z\rho\,\,\bm{\hat \rho} &\qquad\mbox{if}\qquad \rho\leq R\\&\\
\frac{B_0R^2}{\sqrt{2}}\lambda_z\frac{1}{\rho}\bm{\hat\rho} &\qquad\mbox{if}\qquad \rho\geq R,
\end{array}
\right.
\end{equation}
while
\begin{equation}\label{c6}
\int\bm{\alpha\times\beta}\bm\cdot d\bm\sigma=\left\{
\begin{array}{ll}
\sqrt{2}\Phi_0 g_z \,\left(\frac{\rho}{R}\right)^2 &\qquad\mbox{if}\qquad \rho\leq R\\&\\
\sqrt{2}\Phi_0 g_z &\qquad\mbox{if}\qquad \rho\geq R.
\end{array}
\right.
\end{equation}
The interpretation is similar to the case of the magnetic monopole inside the solenoid, but the absence of a Dirac string implies a 
finite charge over the whole space, where the interaction between $\bm\alpha$ and $\bm\beta$ polarizes the whole space and generates 
a charge that behaves in the same fashion as the electric field of a charged linear wire.

\paragraph{\underline{Magnetically charged wire crossing the electrically charged spinning spherical shell}}
In this case, the magnetically charged wire crosses the spherical shell along the spinning axis, through infinitesimal holes. The vector potential $\bm\alpha$ is given in (\ref{r5}) and vector potential $\bm\beta$ is given in (\ref{c2}). Therefore, we have
\begin{equation}\label{c7}
\bm{\alpha\times\beta}=\left\{
\begin{array}{ll}
\sqrt{2}\alpha_0\lambda_z\frac{\rho}{R}\,\,\bm{\hat \rho} &\qquad\mbox{if}\qquad \sqrt{\rho^2+z^2}\leq R\\&\\
\sqrt{2}\alpha_0 R^2\lambda_z\frac{\rho}{\big(\rho^2+z^2\big)^{3/2}}\bm{\hat \rho} &\qquad\mbox{if}\qquad \sqrt{\rho^2+z^2}\geq R,
\end{array}
\right.
\end{equation}
where $\rho=r\sin\theta$ has been used. In a region far from the wire, the interaction term goes with $\rho^{-1}$, and behaves as (\ref{c5}), and in the region inside the spherical shell the behavior is also similar in both of the cases, as expected. On the other hand, 
\begin{equation}\label{c8}
\int\bm{\alpha\times\beta}\bm\cdot d\bm\sigma=\left\{
\begin{array}{ll}
\frac{2\pi\sqrt{2}\alpha_0}{R}g_z\rho^2\,\bm{\hat\rho} &\qquad\mbox{if}\qquad \sqrt{\rho^2+z^2}\leq R\\&\\
2\pi\sqrt{2}\alpha_0 R^2\lambda_z \frac{L}{\sqrt{L^2+\rho^2}}&\qquad\mbox{if}\qquad \sqrt{\rho^2+z^2}\geq R,
\end{array}
\right.
\end{equation}
where the surface integral is integrated for $z\in(-L/2,\,L/2)$ and gives a constant value in the limit of infinite $L$, in agreement with
the previous result (\ref{c6}).
\section{\sc Charge quantization\label{Q}}

Let us consider a construction with a magnetic charge and an electric charge. The spherically symmetric magnetic quaternion monopole is
\begin{equation}\label{q1}
\bm{\mathcal{B}}=\Big(g_0+g_1j\Big)\frac{\bm r}{r^3},
\end{equation}
where $\overline{\bm{\mathcal{B}}}=-\bm{\mathcal{B}}.\,$ At a distance $\,z_0\,$ along the $\,z\,$ axis, a  quaternion imaginary electric charge
$q=q_0+q_1j$ generates a quaternionic electric field $\,\bm{\mathcal{E}}\,$. According to (\ref{m18}), these fields generate a magnetic momentum over each point of the space that 
circularly flows around the $z$ axis and allots an angular momentum $\,\bm{\mathcal{L}}\,$ along this very direction. Summing up the whole space $\,V\,$ we have
\begin{equation}\label{q2}
\bm{\mathcal{L}} =\int_V \bm{r\times\mathcal{P}}_F\,dv,
\end{equation}
where the angular momentum of the fields is
\begin{equation}\label{q3}
\bm{\mathcal{P}}_F\,=\,\frac{1}{8\pi c}\Big(\bm{\mathcal{B}\times\mathcal{E}}\,-\,\bm{\mathcal{E}\times\mathcal{B}}\Big)\,=\,
\,\frac{1}{8\pi c}\Big(2\,\bm{\mathcal{B}_0\times\mathcal{E}_0}\;+\;\bm{\mathcal{E}_1\times\overline{\mathcal{B}}_1}
\;+\;\bm{\overline{\mathcal{E}}_1\times\mathcal{B}_1}\Big),
\end{equation}
where $\,\bm{\mathcal{E}}=\bm{\mathcal{E}}_0+\bm{\mathcal{E}}_1j\,$ and $\,\bm{\mathcal{B}}=\bm{\mathcal{B}}_0+\bm{\mathcal{B}}_1j.\,$ In order to evaluate the angular momentum, let us use (\ref{q2}) in (\ref{q3}) and entertain the first term. 
Using the identity
\begin{equation}\label{q4}
\bm{r\times\mathcal{B}_0\times\mathcal{E}_0}\,=\big(\bm{r\cdot\mathcal{E}}_0\big)\bm{\mathcal{B}_0}-\big(\bm{r\cdot\mathcal{B}}_0\big)\bm{\mathcal{E}_0},
\end{equation}
the definition $\,\bm r=(x_1,\,x_2,\,x_3)\,$ and (\ref{q1}), we decompose the integral into vector components, so that
\begin{equation}\label{q5}
\int_V\Big(\bm{r\times\mathcal{B}_0\times\mathcal{E}_0}\Big)_k\,dv\;=\;g_0\int_V\left[\big(\bm{r\cdot\mathcal{E}}_0\big)\frac{x_k}{r^3}-\frac{1}{r}(\mathcal{E}_0)_k\right]dv\,=\,
-g_0\int_V\bm{\mathcal{E}_0\cdot\nabla}\left(\frac{x_k}{r}\right)dv.
\end{equation}
Integrating by parts, and using the divergence theorem in the second term of (\ref{q5}), we obtain
\begin{equation}\label{q6}
g_0\int_V\left[\frac{x_k}{r}\,\bm{\nabla\cdot\mathcal{E}}_0-\bm{\nabla\cdot}\left(\bm{\mathcal{E}}_0\frac{x_k}{r}\right)\right]dv\,=\, 
g_0\left[4\pi\int_V\frac{x_k}{r}\mathscr{P}_0 dv-\oint\frac{x_k}{r}\bm{\mathcal{E}_0\cdot}d\bm\sigma\right],
\end{equation}
where $\,\bm{\nabla\cdot\mathcal{E}_0}=4\pi\,\mathscr{P}_0\,$ has been used. An identical calculation can be performed in all the terms of (\ref{q3}), and thus the total angular momentum is
\begin{equation}\label{q7}
\bm{\mathcal{L}}\,=\,\frac{1}{c}
\;\int_V\left(g_0 \mathscr{P}_0+\frac{\bar g_1 \mathscr{P}_1+g_1\bar{\mathscr{P}}_1}{2}\right)\bm{\hat r}dv\,-\,
\frac{1}{4\pi c}\oint\frac{\bm{ r}}{r}\left(g_0\bm{\mathcal E}_0+\frac{\bar g_1\bm{\mathcal E}_1+g_1\overline{\bm{\mathcal E}}_1}{2}\right)\bm\cdot d\bm\sigma.
\end{equation}
Remembering that 
\begin{equation}\label{q07}
 \bm{\mathcal{E}}=\big(q_0+q_1j\big)\frac{\bm r-\bm z_0}{|\bm r-\bm z_0|^3},
\end{equation}
the surface integral goes to zero at $r=0$. Furthermore, calculating the volume integral for the electric charge density
\begin{equation}\label{q8}
\mathscr{P}\,=\,\mathscr{P}_0+\mathscr{P}_1 j\,=\,\delta\big(\bm r-\bm z_0\big)\big(q_0+q_1 j\big).
\end{equation}
we obtain
\begin{equation}\label{q9}
\mathcal L_z=\frac{1}{c}\left(g_0 q_0+\frac{\bar g_1 q_1+g_1\bar q_1}{2}\right).
\end{equation}
Imposing $\mathcal L_z =L_z$, where $L_z=(n\hbar)/2$ is the quantum angular momentum, we obtain 
\begin{equation}\label{q10}
 g_0 q_0+\frac{\bar g_1 q_1+g_1\bar q_1}{2}=n\frac{\hbar c}{2}, \qquad\mbox{where}\qquad n\in\mathbb{Z}.
\end{equation}
This quantization rule for quaternionic charges  recovers the Dirac quantization of the electric and magnetic charges for $\;g_1=q_1=0.\;$ As a consequence, we generalized the Dirac
quantization and additionally $\mathbb{H}$ED overcame another consistency test. Let us confirm this result using a the particular case of (\ref{m13}) where
\begin{equation}\label{q11}
 \bm{\mathcal{F}}=\frac{1}{2c}\big(\bm{\mathcal{E}\times\mathcal{J}_M}-\bm{\mathcal{J}_M\times\mathcal{E}}\big)=
 \frac{1}{c}\left(\bm{\mathcal{E}_0\times\mathcal{J}_{M0}}+\frac{\bm{\overline{\mathcal{J}}_{M1}\times\mathcal{E}_1+\mathcal{J}_{M1}\times\overline{\mathcal{E}}_1}}{2}\right),
\end{equation}
with $\,\bm{\mathcal{E}}=\bm{\mathcal{E}}_0+\bm{\mathcal{E}}_1j\,$ and $\,\bm{\mathcal{J}}_M=\bm{\mathcal{J}}_{M0}+\bm{\mathcal{J}}_{M1}j.$ 
The electric field is constant over the $\bm z$ direction while the
current is generated by a monopole of charge $\,g=g_0+g_1j\,$ and mass $m$ that performs a circular motion of constant linear velocity $\bm u$ along the $\bm\phi$ direction, so that 
$\bm{\mathcal{J}}_M=g\bm u$ and the centripetal force equals the electrodynamic force (\ref{q11}), namely
\begin{equation}\label{q12}
\frac{mu}{R}=\frac{1}{c}\left(\mathcal{E}_0g_0+\frac{g_1\bar{\mathcal{E}}_1+\bar g_1\mathcal{E}_1}{2}\right).
\end{equation}
The norm of a uniform electric field can be written in terms of the charge $q$ over the plates 
\begin{equation}\label{q13}
 \mathcal{E}=\frac{4\pi q}{S}=\frac{4q}{R^2},
\end{equation}
where $S$ is the area of the circular section of the plate that contains the circular motion. The left hand side of (\ref{q13}) goes to $mRu$, and is identified to the norm of the angular 
momentum. Therefore, the quatization rule of Landau for such a system, namely
\begin{equation}\label{q14}
 L_z=2n\hbar\qquad\mbox{for}\qquad n\in\mathbb{Z},
\end{equation}
allots back (\ref{q10}), and thus the quantization methods agree. We point out that the above quantization procedure cannot be immediately
applied to the complex dynamics given in Section \ref{tid}, because the momentum $\bm{\mathcal{Q}}_F$ of the imaginary dynamics is identically zero for the monopole fields
(\ref{q1}) and (\ref{q07}), and thus different fields need to be researched. Additionally, there is not an available result quantum angular momentum 
$\bm{\mathcal{K}}=\bm{r\times\mathcal{Q}}_F$  from quaternionic quantum mechanics ($\mathbb{H}$QM). It is however an interesting topic for future research.

\section{\sc Conclusion\label{Co}}
In this article we constructed an electromagnetic theory from pure imaginary quaternionic fields ($\mathbb{H}$ED). The existence of quaternionic magnetic fields
discovered in $\mathbb{H}$QM \cite{Giardino:2019xwm} inspired this research. These quaternionic fields are naturally endowed with magnetic monopoles
and the formalism developed in this article accurately generalizes the real electrodynamics. $\mathbb{H}$QM naturally contains a non-linear
feature concerning the self-interaction of magnetic vector potential. This self-interaction dynamically generate the magnetic charges whose  existence cannot be ascertained without the quaternion structure. Therefore, the quaternionic formalism is a natural way to express such self-interaction phenomenon, despite it is not the unique proposal \cite{Nadi:2019nmi}

There are many directions for future research and the entire electrodynamic theory may be rewritten, and new structures may appear. The physical role of the quaternionic degrees of freedom of the charges and fields is also a matter for future
considerations, and this issue is naturally connected to the physical nature of fields and charges.
The formulation of a quaternionic quantum electrodynamics ($\mathbb{H}$QED) is another matter which new and exciting results may come from,
including the comparison between quaternionic magnetic monopoles and the non-abelian 't Hooft-Polyakov monopole \cite{Shnir:2005xx} in gauge theories.  Additional interesting directions concern the physically interesting problems, such the existence of dyons, and also mathematically oriented problems, like the formulation of quaternionic eletrodynamics in terms of complex quaternions, which is a suitable framework to describe covariant formalism, and can eventually incorporate the idea of self-interaction described here.

\paragraph*{\bf Acknowledgments}
Sergio Giardino gratefully thanks professor Hector Carri\'on (UFRN) for a discussion concerning the gauge transformation within Section \ref{GT}. 

%
%
%
%

\bibliographystyle{unsrt} 
\bibliography{bib_monopolo1}

\end{document}